\documentclass[10pt]{article}
\usepackage{graphicx}
\usepackage{amsmath}
\usepackage{cite}
\usepackage{amsfonts}
\usepackage{amssymb}
\makeatletter
             \@addtoreset{equation}{section}

             \makeatother

\begin{document}

\title{Convective dispersion without molecular diffusion}
\author{Kevin D. Dorfman and Howard Brenner\thanks{Corresponding 
author. Address: 77
Massachusetts Ave., Room 66-564, Cambridge, MA 02139-4307 USA. Phone:
617-253-6687. Fax: 617-258-8224. Email: 
hbrenner@mit.edu.}\\Department of Chemical Engineering 
\\Massachusetts Insitute of Technology \\Cambridge, MA 02139-4307 USA}
\maketitle
\begin{abstract}
A method-of-moments scheme is invoked to compute the asymptotic, long-time
mean (or composite)\ velocity and dispersivity (effective diffusivity) of a
two-state particle undergoing one-dimensional convective-diffusive motion
accompanied by a reversible linear transition (``chemical reaction'' or
``change in phase'') between these states. The instantaneous state-specific
particle velocity is assumed to depend only upon the instantaneous state of
the particle, and the transition between states is assumed to be governed by
spatially-independent, first-order kinetics. \ Remarkably, even in the absence
of molecular diffusion, the average transport of the ``composite''\ particle
exhibits gaussian diffusive behavior in the long-time limit, owing to the
effectively stochastic nature of the overall transport phenomena induced by
the interstate transition. The asymptotic results obtained are compared with
numerical computations.
\end{abstract}

\begin{tabular}
[c]{rl}%
\textbf{Keywords:} & Homogenization, Brownian motion, macrotransport theory,\\
& generalized Taylor dispersion\\

\textbf{PACS:} & 05.40.Jc, 87.10.+e
\end{tabular}

\section{Introduction}

Two-state models are often invoked to construct simple, easily analyzed
transport models of otherwise overwhelmingly complex processes. \ Examples of
such processes include chromatography \cite{Giddings:55,Balakotaiah:95},
isomeric conversion of proteins in solution and \emph{in vivo }calcium
kinetics \cite{Gitterman:95}, and ratchet-driven molecular motor processes
\cite{Julicher:97}. \ In such models, it is typically assumed that the species
of interest, either a single entity (dispersed in a passive solvent) or a
continuum solute concentration field composed of a dilute collection of such
non-interacting species (again dispersed in a passive solvent), undergo(es) a
microscale, state-specific, convective-diffusive transport process wherein the
transport coefficients depend at time $t$ upon the ``state'' $\left(
a,b\right)  $ of the species at that time. \ The rate of interchange between
states is assumed to be governed by first-order kinetics. \ Since the total
species number density does not change in time, this kinetic process is
equivalent in its consequences to a reversible first-order reaction or phase
transition, $a\leftrightarrow b$, between states $a$ and $b$. \ In almost all
cases of interest, one is not generally concerned with the detailed intrastate
transport of the species in state $a$ or $b$ individually, but rather only
with knowledge of the long-time, asymptotic aggregate transport of the
composite $a$-$b$ species; that is, interest focuses only on the combined
transport occurring in both states $a$ and $b$ for times sufficiently long
such that the interstate reaction has effectively achieved global equilibrium
(although the reaction may not be in local equilibrium at each point in
space). \ Remarkably, this averaged aggregate transport is diffusive, even in
the absence of molecular diffusion. \

	Many two-state models only feature the explicit incorporation 
of convection and a ``reaction," without explicitly accounting for 
intrastate molecular diffusion.  Consequently, it is tempting to say 
that these models constitute examples of ``convective dispersion 
without molecular diffusion."  However, the interstate transfer 
reaction often corresponds to a simple model for the solute's {\em 
molecular diffusion,} allowing it to sample two different velocity 
fields prevailing at different points in space, say, in two different 
phases.  In an exactly-posed model \cite{Aris:59}, molecular 
diffusion transports the solute to the interphase boundary, whereupon 
an interstate mass transfer ``reaction" occurs.  Without the explicit 
incorporation of diffusion, a simplified (or rate limited) 
one-dimensional model \cite{Giddings:55,Balakotaiah:95} allows the 
solute to ``jump" between phases in a kinetic manner.  The 
dispersivity computed from such a ``nondiffusive" model 
\cite{Balakotaiah:95}, which possesses a mathematical structure 
similar to the generic problem to be considered here, does indeed 
furnish a convective dispersivity which is explicitly independent of 
the molecular diffusivity.  From a physical viewpoint, though, the 
convective dispersion may still be attributed to molecular diffusion, 
but at a finer-scale level of description than is represented by the 
coarse-scale transport equations adopted in our model.  As such, it 
is not entirely correct to say that existing two-state chromatography 
models satisfy our claim of ``convective dispersion without molecular 
diffusion."

	Rather, our claim more closely corresponds to the case where 
the dispersion arising from the interstate reaction does not occur as 
a consequence of molecular diffusion involving solute molecules which 
randomly shuttle back and forth between two different velocity fields (phases) 
located at different points in space.  To see how such a scenario 
arises, consider the examples of molecular motors and denaturing of 
double-stranded DNA.  In the case of molecular motors 
\cite{Julicher:97}, the species can exist in either a charged or 
uncharged state.  When charged, the particle undergoes convective 
transport due to its interaction with a charged potential field, the 
latter referred to as the ``track."  During the course of its 
convective motion, the particle may also interact biophysically with 
the surrounding medium (say, by incorporating ATP).  The latter 
interactions neutralize the particle and arrest its convective 
motion.\footnote{In more realistic models, the tracks possess a 
spatially-periodic, ratchet-like charge distribution.  While modeling 
periodic systems necessitates a more elaborate moment-matching scheme 
than the one employed here, the existing macrotransport scheme 
\cite{Brenner:93} for single-state systems is readily extensible to 
two-states. The resulting calculations are more involved, but the 
final result reveals, as would be expected, that a similar 
contribution to the convective dispersion exists in the periodic 
case.} The overall biophysical process is cyclical, so the particle 
may again become uncharged, enabling the transition between charged 
and uncharged states to be modeled as a reversible chemical reaction. 
Though the particle itself also undergoes Brownian motion, the 
transitions between the charged and uncharged states depend only upon 
the chemical interactions of the particle with the surrounding 
medium, not the spatial location of the particle.

Another example of a non-diffusive change in state involves the 
partial denaturing of double-stranded DNA.  In such circumstances, 
the two associated strands become locally separated over a short 
range of base pairs, owing to an elevated temperature (or an 
equivalent concentration of denaturant) exceeding the local melting 
temperature.  The native-denatured transition is primarily attributed 
to a chemical interaction between the base pairs (and the denaturant, 
if present), rather than to molecular diffusion. The native and 
denatured states possess distinctly different electrophoretic 
mobilities, and it is these differences which have been exploited in 
the context of numerous electrophoretic schemes \cite{new1,new2} for 
the detection of disease-causing mutations.  Inasmuch as the mutation 
detection relies upon interpreting the chromatographic bands, it 
would be useful to have a greater understanding of the role of 
fluctuations between the states (which could be related to, say, 
fluctuations in the effective temperature of the medium) upon the 
overall chromatographic process.  In both the motor and denatured DNA 
cases, the change in the solute state, and thus the solute velocity, 
arises from a chemical interaction with the medium, not from a 
conventional molecular diffusion process transporting the solute to a 
new point in space.  Consequently, the convective dispersion process 
arising here would persist in the absence of molecular diffusion.

As a prelude to a much more general analysis, the main point of this paper is
illustrated by considering the following elementary non-diffusive system:
\begin{align}
\frac{\partial P_{a}}{\partial t}+U\frac{\partial P_{a}}{\partial x}+k\left(
P_{a}-P_{b}\right)   & =\frac{1}{2}\delta\left(  x\right)  \delta\left(
t\right)  ,\label{Psimple1}\\
\frac{\partial P_{b}}{\partial t}-U\frac{\partial P_{b}}{\partial x}+k\left(
P_{b}-P_{a}\right)   & =\frac{1}{2}\delta\left(  x\right)  \delta\left(
t\right)  ,\label{Psimple2}%
\end{align}
where $P_{\alpha}\left(  x,t\right)  $ is the conditional probability density
of state $\alpha$, $U$ is the species velocity, $k$ is the interstate reaction
rate, and $\delta$ is the Dirac delta function. \ Let $\hat{P}_{\alpha}\left(
q,s\right)  $ denote the Fourier-Laplace transform of $P_{\alpha}\left(
x,t\right)  $, with $q$ the Fourier variable and $s$ the Laplace variable.
\ Transforming eqs.\ (\ref{Psimple1})-(\ref{Psimple2}) into Fourier-Laplace
space furnishes the following coupled set of algebraic equations:
\begin{equation}
s\hat{P}_{a}+\left(  iqU\right)  \hat{P}_{a}+k(\hat{P}_{a}-\hat{P}_{b}%
)=\frac{1}{2},\text{\quad}s\hat{P}_{b}-\left(  iqU\right)  \hat{P}%
_{b}+k  (\hat{P}_{b}-\hat{P}_{a})  =\frac{1}{2}.
\end{equation}
Upon solving for $\hat{P}_{a}$ and $\hat{P}_{b}$ and forming their sum, the
Fourier-Laplace transform of the total probability density adopts the form
\begin{equation}
\hat{P}=\hat{P}_{a}+\hat{P}_{b}=\left(  s+\frac{q^{2}U^{2}}{s+2k}\right)
^{-1}.
\end{equation}
For long times $t\gg2k$, corresponding here to $s\ll2k$, this becomes
\begin{equation}
\hat{P}\approx\left(  s+\frac{q^{2}U^{2}}{2k}\right)  ^{-1}.
\end{equation}
The latter expression is readily inverted to yield the gaussian distribution,
\begin{equation}
P\approx\left(  4\pi\bar{D}^{\ast}t\right)  ^{-1/2}\exp\left[ - \frac{x^{2}%
}{4\bar{D}^{\ast}t}\right]  ,\label{diffusioneqn}%
\end{equation}
with an effective diffusion (dispersion) coefficient
\begin{equation}
\bar{D}^{\ast}=\frac{U^{2}}{2k}.\label{Dsimple}%
\end{equation}
Moreover, the mean position of the composite particle remains at the initial
position $x=0$. \ The latter property is equivalent in its consequences to a
zero composite particle velocity, $\bar{U}^{\ast}=0$
[compare\ eq.\ (\ref{gaussian})].

In conventional long-time average ``macrotransport'' analyses
\cite{Brenner:93} of this type, the dispersivity $\bar{D}^{\ast}$ arises via
the stochastic sampling of local velocity inhomogeneities by solute diffusion
in the continuous local space. For example, in the classic case of Taylor
dispersion \cite{Taylor:53}, quantifying the global solute transport occurring
in a Poiseuille flow within a long circular cylindrical tube, radial diffusion
across the streamlines enables a solute molecule being transported axially by
the Poiseuille flow to sample the fluid's transversely parabolic axial
velocity profile innumerable times as it moves downstream, eventually
attaining a stationary mean velocity $\bar{U}^{\ast}$ and dispersivity
$\bar{D}^{\ast}$. \ In contrast, the source of the dispersion in our two-state
problem resides in the interstate reaction, which allows the composite $a$-$b$
species to sample the discrete ``state space.'' \ Nevertheless, dispersion
arising from the sampling of different state-specific velocities is completely
analogous to the dispersion which arises from so-called ``local-space''
velocity inhomogeneities, e.g.\ Poiseuille flow, in classical macrotransport
theory \cite{Brenner:93}. \ The stochastic foundation of chemical reactions is
well established \cite{McQuarrie:67}, inasmuch as the reaction rate $k$ may be
interpreted as reflecting the probability of a reaction occurring in the
interval between times $t$ and $t+\delta t$. Moreover, whereas chemical
reactions are considered to constitute purely deterministic processes in the
full position-momentum space (``local space''), the process is nevertheless
stochastic in the reduced space (``global space'') employed in conventional
kinetic models \cite{Gillespie:77}. \ Likewise, the comparable reduction from
transport in the distinct states $a$ and $b$ to transport in the composite
$a$-$b$ state results in a stochastic (diffusive)\ composite transport process.

In the present contribution, we investigate more thoroughly this surprising
property of two-state systems. \ This is effected by computing, via the theory
of macrotransport processes \cite{Brenner:93}, the asymptotic, long-time mean
velocity, $\bar{U}^{\ast}$, and dispersivity, $\bar{D}^{\ast}$, of a more
general two-state system than that defined by eqs.\ (\ref{Psimple1}%
)-(\ref{Psimple2}). \ This scheme may be likened to a multiple-time scale
analysis \cite{Pagitsas:86} of the phenomena, where the short-time, two-state
behavior ultimately serves to determine the long-time, state-independent
temporal behavior of the system as a whole. \ In a very different,
non-reactive context than that considered here, moment-matching concepts have
been used \cite{Aris:59,Haber:93} to consider systems where the different
``states'' correspond to the different physical phases (e.g.\ solid and
liquid) through which a solute molecule can be transported at different
velocities. \ In contrast with the present analysis, the latter analyses
employed molecular diffusion to transport the solute to the interface across
which the interstate mass transfer (change in state) occurs. \ Subsequently,
Iosilevskii and Brenner \cite{Iosilevskii:95} used eigenfunction expansions of
the moments to develop a general macrotransport scheme for the analysis of
reactive mixtures in an incompressible solvent flow, where molecular diffusion
again played an important role in enabling the solute to sample the local
space. \ Balakotaiah and Chang \cite{Balakotaiah:95} used center-manifold
techniques to examine problems in chromatography theory which are similar to
those considered here, albeit with an implicit dependence upon 
molecular diffusion.
  \ The results of the latter correspond to a special
case of the subsequent analysis, which, we believe, employs a more
straightforward technique than is manifest in center manifold
theory.\footnote{It should be noted that the moment-matching scheme employed
here is only valid for linear (first-order)\ reactions, whereas the
center-manifold technique may be employed to analyze nonlinear transport
processes as well \cite{Balakotaiah:95}.}

\section{Generalized Problem Statement}

Consider the conditional probability density $P_{\alpha}(x,t\,|\,\phi_{\alpha
})$ ($\alpha=a,b$) that a non-interacting collection of particles exists in
state $\alpha$ and are present at position $x$ at time $t$, given their
initial impulsive introduction into the unbounded system $\left(
-\infty<x<\infty\right)  $ at $x=0$ and time $t=0$, with $\phi_{\alpha}$ the
fraction of the particles initially in state $\alpha$. \ The two $\phi
_{\alpha}$ are not independent parameters, since $\phi_{a}+\phi_{b}=1$. \ If
an effective or composite $a$-$b$ ``particle'' description is to exist, the
final results for $\bar{U}^{\ast}$ and $\bar{D}^{\ast}$ quantifying the
spatio-temporal transport of this composite ``particle''\ must (and will)
prove to be independent of the arbitrary choice of labels $a$ and $b$.
\ Moreover, the macrotransport parameters $\bar{U}^{\ast}$ and $\bar{D}^{\ast
}$ will further prove to be independent of the particle's initial arbitrary
position $x=0$, as well as of the initial state fractions $\phi_{\alpha}$.

The spatio-temporal evolution of the conditional probability densities,
$P_{a}\equiv P_{a}\left(  x,t\,|\,\phi_{a}\right)  $ and $P_{b}\equiv
P_{b}\left(  x,t\,|\,\phi_{b}\right)  $, are governed by the coupled set of
convection-diffusion-reaction equations,
\begin{equation}
\frac{\partial P_{\alpha}}{\partial t}+U_{\alpha}\frac{\partial P_{\alpha}%
}{\partial x}-D_{\alpha}\frac{\partial^{2}P_{\alpha}}{\partial x^{2}}+k\left(
K_{\alpha}P_{\alpha}-K_{\beta}P_{\beta}\right)  =\phi_{\alpha}\delta
(t)\delta(x),\label{Paeqn}%
\end{equation}
valid for $\alpha=\left(  a,b\right)  $ $\left(  \alpha\neq\beta\right)  $,
with $\delta$ the Dirac delta function. \ Here, the constants $U_{\alpha}$ and
$D_{\alpha}$ represent the state-specific velocity and molecular diffusivity
of the particle. \ \ It is assumed that the latter parameters depend solely
upon the state $\left(  a,b\right)  $ of the particle, being independent of
position $x$, as well as of time $t$. \ \ While it is possible in the interest
of simplicity to set the state diffusivities $D_{\alpha}$ to zero at the
outset, we retain these terms in order to assess what will prove to be their
relatively straightforward impact upon the averaged long-time transport
process [cf.\ eqs.\ (\ref{Dstar})-(\ref{Dm})]. \

In the above, the transition between states has been assumed to occur via
first-order kinetics, with $kK_{a}$ representing the transition rate from
state $a$ to $b$, and $kK_{b}$ the transition rate from $b$ to $a$, with all
reaction parameters being positive: $(k,K_{a},K_{b})>0$. \ We employ the trio
of parameters $k$, $K_{a}$ and $K_{b}$ (of which only two are independent, say
$kK_{a}$ and $K\equiv K_{b}/K_{a}$), rather than a pair of reaction rates,
say, $k_{a}=kK_{a}$ and $k_{b}=kK_{b}$, so as to permit a clear distinction to
be made between the \emph{kinetic} rate of interstate transfer, embodied in
$k$, and the \emph{stationary} (equilibrium)\ partitioning of states, embodied
in $K_{b}/K_{a}$. Moreover, this egalitarian notational choice serves
simultaneously to emphasize that the final results are independent from the
arbitrary choice of labels $a$ and $b$. \

This coupled set of partial differential equations can be solved, in
principle, subject to the attenuation of the conditional probability densities
and fluxes\footnote{When the $D_{\alpha}$ are non-zero, the $P_{\alpha}$ are
expected to decay exponentially fast as $\left|  x\right|  \rightarrow\infty$,
whence the flux attenuation condition in eq.\ (\ref{infdecay}) will be
automatically satisfied. For the case where the $D_{\alpha}$ are identically
zero, the governing equation (\ref{Paeqn}) is hyperbolic, whereupon the
probability density is identically zero for all times as $\left|  x\right|
\rightarrow\infty$.} at infinity for all times $t>0$:
\begin{equation}
P_{\alpha},J_{\alpha}\rightarrow0\quad\text{as\quad}\left|  x\right|
\rightarrow\infty.\label{infdecay}%
\end{equation}
Summing eqs.\ (\ref{Paeqn}) over $a$ and $b$, and subsequently integrating
over the interval $\left(  -\infty,\infty\right)  $ furnishes the particle
conservation relation,
\begin{equation}
\int_{-\infty}^{\infty}\left(  P_{a}+P_{b}\right)  \,dx=1\quad\left(
\ t>0\right)  ,\label{Pnormal}%
\end{equation}
which reflects the fact that each of the particles originally introduced into
the system at time $t=0$ are conserved for all time, independently of the
initial distributions, $\phi_{a}$ and $\phi_{b}$.

\section{Macrotransport Analysis\label{MacrotransportAnalysis}}

In Appendix A, we use standard moment-matching techniques \cite{Brenner:93} to
analyze the generalized problem for asymptotically long times, namely those
satisfying the inequality
\begin{equation}
t\gg\left[  k\left(  K_{a}+K_{b}\right)  \right]  ^{-1}.\label{longtime}%
\end{equation}
Explicitly, the macrotransport parameters $\bar{U}^{\ast}$ and $\bar{D}^{\ast
}$ are computed by matching the asymptotic rates of change of the total
moments of the microscale probability density,
\begin{equation}
M_{m}\left(  t\,|\,\phi_{\alpha}\right)  \overset{\text{def.}}{=}\int
_{-\infty}^{\infty}x^{m}\left(  P_{a}+P_{b}\right)  dx\quad\left(
m=0,1,2,\ldots\right)  ,\label{Mtotal}%
\end{equation}
against the corresponding moments,
\begin{equation}
\bar{M}_{m}\left(  t\right)  \overset{\text{def.}}{=}\int_{-\infty}^{\infty
}x^{m}\bar{P}\left(  x,t\right)  dx\quad\left(  m=0,1,2,\ldots\right)  ,
\end{equation}
of the macrotransport equation,
\begin{equation}
\frac{\partial\bar{P}}{\partial t}+\bar{U}^{\ast}\frac{\partial\bar{P}%
}{\partial x}-\bar{D}^{\ast}\frac{\partial^{2}\bar{P}}{\partial x^{2}}%
=\delta\left(  x\right)  \delta\left(  t\right)  .\label{macroeqn}%
\end{equation}
In the latter, $\bar{P}$ is the so-called macrotransport probability density
of the composite $a$-$b$ particle, governing the long-time asymptotic
evolution of the true probability density, $P=P_{a}+P_{b}$. \ Solving equation
(\ref{macroeqn}) for $\bar{P}$ furnishes the gaussian distribution,
\begin{equation}
\bar{P}=\left(  4\pi\bar{D}^{\ast}t\right)  ^{-1/2}\exp\left[ - \frac{\left(
x-\bar{U}^{\ast}t\right)  ^{2}}{4\bar{D}^{\ast}t}\right]  .\label{gaussian}%
\end{equation}

With respect to the parameters appearing in eq.\ (\ref{Paeqn}), application of
the moment-matching scheme furnishes the composite Lagrangian velocity,
\begin{equation}
\bar{U}^{\ast}=\frac{K_{b}U_{a}+K_{a}U_{b}}{K_{a}+K_{b}}\equiv\frac
{KU_{a}+U_{b}}{1+K},\label{Ustar}%
\end{equation}
where the parameter
\begin{equation}
K\overset{\text{def.}}{=}\frac{K_{b}}{K_{a}}\label{Kdef}%
\end{equation}
constitutes an equilibrium constant, representing the condition eventually
achieved in eq.\ (\ref{Paeqn}) when $K_{a}P_{a}=K_{b}P_{b}$, i.e.\ when the
``forward''\ and ``backward'' rates at which the states change are in balance,
at least in the global sense embodied in eq.\ (\ref{Painf}). Consequently, the
composite particle velocity (\ref{Ustar}) represents a weighted average of the
respective state-specific velocities, $U_{\alpha}$, with the weighting
corresponding to the stationary partitioning of states at long
times.\footnote{Substitution of the parameters appearing in our initial
example (\ref{Psimple1})-(\ref{Psimple2}) into eq.\ (\ref{Ustar}) confirms
that $\bar{U}^{\ast}=0$ in that case.} \ Albeit in a very different context,
this result agrees with the comparable two-phase result of Aris (sans areal
factors) \cite{Aris:59}, while also reducing to the results of Balakotaiah and
Chang \cite{Balakotaiah:95} upon setting one of the two state velocities to
zero. \ Importantly, the mean velocity depends solely upon the stationary
partitioning of the particle states, $K$, rather than being dependent upon the
kinetic constant, $k$, governing the rate at which the particle transits
between states before achieving this stationary partitioning of states.

The moment-matching scheme also furnishes the composite particle
dispersivity,
\begin{equation}
\bar{D}^{\ast}=\bar{D}^{M}+\bar{D}^{C},\label{Dstar}%
\end{equation}
where
\begin{equation}
\bar{D}^{M}=\frac{K_{b}D_{a}+K_{a}D_{b}}{K_{a}+K_{b}}\equiv\frac{KD_{a}+D_{b}%
}{1+K},\label{Dm}%
\end{equation}
represents the molecular or ``Aris'' contribution to the total dispersivity,
and
\begin{equation}
\bar{D}^{C}=\frac{K_{a}K_{b}}{\left(  K_{a}+K_{b}\right)  ^{3}}\frac{\left(
U_{a}-U_{b}\right)  ^{2}}{k}\equiv\frac{K}{\left(  1+K\right)  ^{3}}%
\frac{\left(  U_{a}-U_{b}\right)  ^{2}}{kK_{a}}\label{Dc}%
\end{equation}
constitutes the corresponding convective or ``Taylor'' dispersion. \ Of
course, eqs.\ (\ref{Dstar})-(\ref{Dc}) reduce appropriately to (\ref{Dsimple})
in the special state-specific circumstances quantifying the latter. \ It is
readily confirmed that $\bar{D}^{C}$ is positive semi-definite, vanishing only
when: (i) the state velocities are identical in magnitude and direction; or
(ii) one of the two parameters, $K_{\alpha}$, vanishes. \ The latter
corresponds to the trivial transport process where, for long-times, the
particle exists in but a single state. \ The functional form of eq.\ (\ref{Dc}%
) is equivalent (to within a prefactor) to that derived by Aris \cite{Aris:59}%
. \ Balakotaiah and Chang's result \cite{Balakotaiah:95} is again recovered by
setting one state velocity to zero.

At the present juncture, it is informative to compare the convective
dispersivity, eq.\ (\ref{Dc}), with the classical result
\cite{Taylor:53,Aris:56} for the convective contribution to the Taylor-Aris
dispersion of a solute entrained in a Poiseuille flow moving at mean velocity
$\bar{v}$ through a cylindrical tube of radius $R$, namely
\begin{equation}
\bar{D}^{C}=\frac{1}{48}\frac{\bar{v}^{2}R^{2}}{D},\label{Taylor}%
\end{equation}
where $D$ is the molecular diffusivity of the solute. \ The quadratic
dependence upon the velocity ``inhomogeneity,'' $\left(  U_{a}-U_{b}\right)
^{2}$, appearing in (\ref{Dc}) is analogous to the functional dependence,
$\bar{v}^{2}$, appearing in (\ref{Taylor}). \ Moreover, the factor $k^{-1}$
appearing in (\ref{Dc}) possesses an interpretation identical to the factor
$R^{2}/D$ appearing in (\ref{Taylor}). \ Explicitly, each quantity
respectively represents the characteristic time required for sampling the
local space, either the ``state'' space in the present reactive problem or the
physical space embodied in the tube's cross-sectional area, $\pi R^{2}$.

In contrast with $\bar{U}^{\ast}$, which is independent of the kinetic
properties embodied in $k$, the convective dispersivity $\bar{D}^{C}$ depends
upon both the particle's equilibrium and kinetic properties, respectively
embodied in the parameters $K$ and $k$. \ The physical basis for this kinetic
dependence may be rationalized by considering the transport of two particles,
each occupying identical positions $x$ and states $\alpha$ at time $t$. \ Were
both particles to remain in state $\alpha$, deviations in their relative mean
positions would arise only from molecular diffusion, a phenomenon which is
accounted for by the molecular contribution, $\bar{D}^{M}$, to the
dispersivity, $\bar{D}^{\ast}$. \ In contrast, were the first particle to
switch to state $\beta$ while the second remained in state $\alpha$, these two
particles would tend to separate (``spread'') due to the relative difference
in their state-specific particle velocities, $\left|  U_{\alpha}-U_{\beta
}\right|  $. \ Consider first the case where the reaction rate is rapid
relative to the rate of convective spreading, i.e.$\ $ $k\gg\left|  U_{\alpha
}-U_{\beta}\right|  /\Delta x$, with $\Delta x$ a characteristic separation
distance. [The combination $k\left(  \Delta x\right)  ^{2}$ represents the
coupling of interstate (local) transport to global transport, which is
identical to the physical interpretation underlying $\bar{D}^{C}$.] \ For fast
reactions, the likelihood is great that the particles will return to an
identical state (either $\alpha$ or $\beta$), and hence identical velocity, in
a brief period of time, thereby minimizing their ``spreading.'' \ In contrast,
for slow reactions, $k\ll\left|  U_{\alpha}-U_{\beta}\right|  /\Delta x$,
there exists a high probability that the particles will travel significant
distances before returning to identical states (and identical velocities),
whereupon the ``spreading'' caused by the interstate transfer is expected to
be much larger than in the fast reaction case. \ The appearance in $\bar
{D}^{C}$ of the square, $\left(  U_{a}-U_{b}\right)  ^{2}$, of the velocity
difference rather than the term $\left|  U_{a}-U_{b}\right|  $ invoked in the
preceding scaling arguments, arises from the fact that the dispersivity
represents an asymptotic measure of the mean-squared deviation of the relative
particle positions, rather than simply their absolute deviation.

An alternative rationalization of the dispersivity $\bar{D}^{\ast}$ is
achieved by considering its role in quantifying the deviations of the
instantaneous solute particle position, $x$, at time $t$ from its composite
position, $\bar{U}^{\ast}t$, which the particle would occupy at time $t$ if it
always moved uniformly at the velocity $\bar{U}^{\ast}$. \ To aid in this
analysis, define the convected coordinate variable,
\begin{equation}
x^{\ast}\overset{\text{def.}}{=}x-\bar{U}^{\ast}t.\label{xstar}%
\end{equation}
Appropriate conversion of the microscale equation (\ref{Paeqn}) into this
altered coordinate format furnishes the following equation governing the
respective $\left(  a,b\right)  $ transport processes:
\begin{equation}
\frac{\partial P_{\alpha}}{\partial t}+\left(  U_{\alpha}-\bar{U}^{\ast
}\right)  \frac{\partial P_{\alpha}}{\partial x^{\ast}}-D_{\alpha}%
\frac{\partial^{2}P_{\alpha}}{\partial x^{\ast^{2}}}+k\left(  K_{\alpha
}P_{\alpha}-K_{\beta}P_{\beta}\right)  =\phi_{\alpha}\delta\left(  x^{\ast
}\right)  \delta\left(  t\right)  ,\label{Pdim1}%
\end{equation}
where, here, $\partial/\partial t=\left(  \partial/\partial t\right)
_{x^{\ast}}$. \ In this new coordinate system, one of the two state-specific
velocities takes place in the $+x^{\ast}$ direction, whereas the other occurs
in the $-x^{\ast}$ direction (except for the trivial case where both state
velocities are equal), since $\bar{U}^{\ast}$ represents an average of the
respective state velocities, $U_{\alpha}$. \ Consequently, the transport
process occurring in the convected $x^{\ast}$ coordinate system may be
envisioned as a biased random walk, where the state-dependent step-sizes are
proportional to the velocity difference, $\left|  U_{\alpha}-\bar{U}^{\ast
}\right|  $, while the probability of taking a step in the $\alpha$-direction
is proportional to $K_{\beta}.$

\section{Comparison with Numerical Results}

Numerical solutions are presented here for several illustrative choices of the
transport and reaction parameters, comparing both the short- and long-time
evolutions of the exact microscale probability density, $P=P_{a}+P_{b}$, with
its asymptotic counterpart, $\bar{P}$, given by eq.\ (\ref{gaussian}). \ For
simplicity, we have eliminated intrastate molecular diffusion effects by
setting $D_{a}=D_{b}=0$ in (\ref{Pdim1}). \ Anticipating that, for long times,
the total solute probability density will be convected at the rate
$\bar{U}^{\ast}$, as in (\ref{Ustar}), we proceed with the formulation
(\ref{Pdim1}) of the microscale transport problem in the convected coordinate
system, $x^{\ast}$. In rendering eq.\ (\ref{Pdim1}) dimensionless, it proves
convenient to abandon the egalitarian notation employed thus far, choosing
state $a$ as the base state.\footnote{Equivalent results may, of course, be
obtained by choosing state $b$ as the base state and interchanging all
subscripts.} \ With dimensionless time and length variables chosen as
\begin{equation}
\tilde{t}=tkK_{a},\quad\tilde{x}^{\ast}=x\left(  \frac{kK_{a}}{U_{a}}\right)
,
\end{equation}
eq.\ (\ref{Pdim1}) adopts the respective dimensionless forms%
\begin{align}
\frac{\partial P_{a}}{\partial\tilde{t}}+\tilde{U}_{a}\frac{\partial P_{a}%
}{\partial\tilde{x}^{\ast}}+\left(  P_{a}-KP_{b}\right)    & =\phi_{a}%
\delta\left(  \tilde{x}^{\ast}\right)  \delta\left(  \tilde{t}\right)
,\label{Pdima}\\
.\frac{\partial P_{b}}{\partial\tilde{t}}+\tilde{U}_{b}\frac{\partial P_{b}%
}{\partial\tilde{x}^{\ast}}-\left(  P_{a}-KP_{b}\right)    & =\left(
1-\phi_{a}\right)  \delta\left(  \tilde{x}^{\ast}\right)  \delta\left(
\tilde{t}\right)  .\label{Pdimb}%
\end{align}
The equilibrium constant $K$ is given by (\ref{Kdef}), whereas the
dimensionless state-specific velocities possess the respective forms
\begin{equation}
\tilde{U}_{a}=\frac{1-\tilde{U}}{1+K},\quad\tilde{U}_{b}=-K\tilde{U}_{a},
\end{equation}
with $\tilde{U}$ the velocity ratio,
\begin{equation}
\tilde{U}\overset{\text{def.}}{=}\frac{U_{b}}{U_{a}}.
\end{equation}
In the convected, dimensionless coordinate system, our asymptotic result
(\ref{gaussian}) adopts the form
\begin{equation}
\bar{P}=\left(  4\pi\tilde{D}^{\ast}t\right)  ^{-1/2}\exp\left[ - \frac{\left(
\tilde{x}^{\ast}\right)  ^{2}}{4\tilde{D}^{\ast}t}\right]  ,
\end{equation}
with $\tilde{D}^{\ast}$, the dimensionless dispersivity,%
\begin{equation}
\tilde{D}^{\ast}=\frac{K\left(  1-\tilde{U}\right)  ^{2}}{\left(  1+K\right)
^{3}}.
\end{equation}

The coupled equation set (\ref{Pdima})-(\ref{Pdimb}) was solved numerically
for various values of the parameters $K$, $\tilde{U}$ and $\phi_{a}$, using
upwind finite differences for the advected terms and forward Euler time
integration. \ Numerical results obtained for short ($\tilde{t}=0.5$) and long
($\tilde{t}=50$) times are depicted in Figs.\ 1 and 2, respectively, along
with the corresponding asymptotic macrotransport results. \ Similar results
(not depicted) were obtained for other choices of the parameters $K,$
$\tilde{U}$ and $\phi_{a}$.

The results shown in Fig.\ 1 demonstrate that, for short-times, the system may
exhibit dramatically different behavior from that displayed at longer times,
depending upon the particular choices made for the parameters $K$ and
$\tilde{U}$ and initial state $\phi_{a}$ of the system. \ In general, the
short-time transport of $a$ and $b$ as a whole is, not unexpectedly, bimodal,
since that species whose velocity exceeds $\bar{U}^{\ast}$ moves in the
$+\tilde{x}^{\ast}$ direction, while its counterpart moves in the $-\tilde
{x}^{\ast}$ direction. \ Consequently, our asymptotic macrotransport analysis,
which is symmetric about the origin $\tilde{x}^{\ast}=0$, invariably proves
highly inaccurate when attempting to capture the short-time behavior,
typically overestimating the width of the distribution. \ In contrast, as
depicted in Fig.\ 2, the
macrotransport description agrees very well with the numerical solutions for
long-times for all choices of the parameters examined.
\ As the species separate, the interstate reaction serves to
``mix'' their probability densities, ultimately giving rise to an
asymptotically gaussian distribution. \

It should be noted that the hyperbolic structure of the non-diffusive
microscale equation (\ref{Pdima})-(\ref{Pdimb}) guarantees that, except for
the special case $\tilde{U}_{a}=\tilde{U}_{b}$, the exact solution of the
microscale equations will be asymmetric with respect to the origin.
\ Explicitly, at time $\tilde{t}^{\ast}$, the maximum, $\tilde{x}_{\max}%
^{\ast}$, and minimum, $\tilde{x}_{\min}^{\ast}$, \ possible spatial positions
with non-zero probability densities are given respectively by $\tilde{x}%
_{\max}^{\ast}=\tilde{U}_{a}\tilde{t}^{\ast}$ and $\tilde{x}_{\min}^{\ast
}=\tilde{U}_{b}\tilde{t}^{\ast}$ (taking $\tilde{U}\geq1$ without any loss of
generality). \ Aside from the aforementioned special case, the exact solution
is expected to be asymmetric, since $\tilde{x}_{\max}^{\ast}\neq\tilde
{x}_{\min}^{\ast}$. \ In contrast, the macrotransport solution is not only
symmetric about the origin, but also predicts a non-zero probability density
for all values of $\tilde{x}^{\ast}$. \ However, this asymmetry-induced
disparity existing between the exact solution and the macrotransport solution
occurs in the tails of the probability density, at which positions the latter
density is already exponentially small, and therefore of the order of the
error incurred in our asymptotic analysis. \ In contrast, the differences
between the exact and the macrotransport solutions are expected to be small in
the ``central region'' for long times, as confirmed by the numerical results.

\section*{Acknowledgments}

This work was supported in part by a Graduate Research Fellowship awarded to
KDD\ by the National Science Foundation. We acknowledge useful discussions
regarding the numerical solutions with Scott D. Phillips of MIT.

\appendix

\section{Moment-Matching Scheme}

Define the state-specific $m^{th}$-order ``local'' moment of the probability
density of state $\alpha=\left(  a,b\right)  $ as
\begin{equation}
P_{\alpha}^{(m)}(t\,|\,\phi_{\alpha})\overset{\text{def.}}{=}\int_{-\infty
}^{\infty}x^{m}P_{\alpha}(x,t\,|\,\phi_{\alpha})dx\quad\left(  m=0,1,2,\ldots
\right)  .\label{state}%
\end{equation}
The latter moments are equivalent to the local space (bounded) variables in
conventional macrotransport theory \cite{Brenner:93}. \ In order to guarantee
that these higher-order moments are finite, it is necessary to strengthen the
attenuation condition (\ref{infdecay}), such that the probability densities
and fluxes decay faster than algebraically, namely
\begin{equation}
\left|  x\right|  ^{m}P_{\alpha}\rightarrow0\quad\text{and\quad}\left|
x\right|  ^{m}\partial P_{\alpha}/\partial x\rightarrow0\quad\text{as\quad
}\left|  x\right|  \rightarrow\infty.\label{Pattenuate}%
\end{equation}
The differential equation governing the state-specific local moment
(\ref{state}) is derived as follows: (i) multiply eq.\ (\ref{Paeqn}) by
$x^{m}$; (ii) integrate over the range $-\infty<x<\infty$; (iii) integrate the
resulting expression by parts; and (iv) subsequently apply the attenuation
conditions (\ref{Pattenuate}). \ This scheme eventually furnishes the relation
(for $\alpha\neq\beta$)
\begin{align}
\frac{dP_{\alpha}^{(m)}}{dt}+k\left[  K_{\alpha}P_{\alpha}^{(m)}-K_{\beta
}P_{\beta}^{(m)}\right]     =& mU_{\alpha}P_{\alpha}^{(m-1)}+m\left(
m-1\right)  D_{\alpha}P_{\alpha}^{\left(  m-2\right)  }+\nonumber\\
& +\ \phi_{\alpha}\delta(t)\delta_{m0},\label{Pastate}%
\end{align}
where $\delta_{m0}$ is the Kronecker delta function (i.e. $\delta_{00}=1$ and
$\delta_{m0}=0$ for $m\neq0$). \ In principle, this equation can be solved
recursively for $P_{\alpha}^{(m)}(t\,|\,\phi_{\alpha})$, although with use of
macrotransport theory it proves unnecessary to do so in order to eventually
calculate $\bar{U}^{\ast}$ and $\bar{D}^{\ast}$.

In a similar manner, the $m^{th}$-order ``total moment'' of the composite
particle represents the sum of the corresponding local moments
[cf.\ eq.\ (\ref{Mtotal})],
\begin{equation}
M_{m}(t\,|\,\phi_{\alpha})=P_{a}^{(m)}+P_{b}^{(m)}\quad\left(  m=0,1,2,\ldots
\right)  .\label{total}%
\end{equation}
Summing eq.\ (\ref{Pastate}) over the two states, $\alpha=\left(  a,b\right)
$, furnishes the following differential equation governing each $M_{m}$:
\begin{align}
\frac{dM_{m}}{dt}  =& m\left[  U_{a}P_{a}^{(m-1)}+U_{b}P_{b}^{(m-1)}\right]
+m(m-1)\left[  D_{a}P_{a}^{(m-2)}+D_{b}P_{b}^{(m-2)}\right]  +\nonumber\\
&   + \ \delta(t)\delta_{m0}.\label{Meqn}%
\end{align}
The macrotransport parameters may then be computed from the following
relationships \cite{Brenner:93}:
\begin{align}
\bar{U}^{\ast} &  =  \lim_{t\rightarrow\infty}\frac{dM_{1}}{dt},\label{Udef}\\
\bar{D}^{\ast} &  = \lim_{t\rightarrow\infty}\frac{d}{dt}\left(  M_{2}%
-M_{1}^{2}\right)  ,\label{Ddef}%
\end{align}

We proceed here to solve eqs.\ (\ref{Pastate}) and (\ref{Meqn}), at least in
the long-time limit (\ref{longtime}). \ This enables us to establish the
asymptotic forms of the moments, $P_{\alpha}^{(m)}$ and $M_{m}$, for
$m=0,1,2$, thereby permitting $\bar{U}^{\ast}$ and $\bar{D}^{\ast}$ to be
calculated via eqs.\ (\ref{Udef})-(\ref{Ddef}). \ Upon setting $m=0$ in
eq.\ (\ref{Meqn}) and integrating with respect to $t$, the zeroth-order total
moment is found to be
\begin{equation}
M_{0}=\left\{
\begin{array}
[c]{l}%
0\quad\left(  t\leq0\right)  ,\\
1\quad\left(  t>0\right)  ,
\end{array}
\right. \label{M0}%
\end{equation}
reflecting the conservation condition (\ref{Pnormal}).

The zeroth-order, state-specific local moment equation is obtained by setting
$m=0$ in eq.\ (\ref{Pastate}), yielding
\begin{equation}
\frac{dP_{\alpha}^{0}}{dt}+k\left(  K_{\alpha}P_{\alpha}^{0}-K_{\beta}%
P_{\beta}^{0}\right)  =\phi_{\alpha}\delta(t)\quad\left(  \alpha\equiv
a,b\right)  .\label{Pa0}%
\end{equation}
The asymptotic solution of this pair of equations, valid for times satisfying
the inequality (\ref{longtime}), is given by
\begin{equation}
P_{\alpha}^{0}\approx P_{\alpha}^{0,\infty}+\text{exp},\quad P_{\alpha
}^{0,\infty}=\frac{K_{\beta}}{K_{a}+K_{b}},\label{Painf}%
\end{equation}
where the symbol ``exp'' denotes position- and initial condition-dependent
functions that are attenuated exponentially rapid in time. \ By forming the
sum $P_{a}^{0,\infty}+P_{b}^{0,\infty}$ from eq.\ (\ref{Painf}), it is seen
that the state-specific solutions (\ref{Painf}) asymptotically satisfy the
normalization condition (\ref{Pnormal}). \ Moreover, for long times, the local
moments $P_{\alpha}^{0,\infty}$ are unconditional (rather than conditional)
probability densities, being independent of both the initial position and the
instantaneous state of the particle at $t=0$, as was to be expected. \

The composite velocity $\bar{U}^{\ast}$ is computed by setting $m=1$ in
eq.\ (\ref{Meqn}), substituting the asymptotic solution (\ref{Painf}) for
$\alpha=\left(  a,b\right)  $ into the resulting differential equation, and
applying eq.\ (\ref{Udef}), so as to obtain the result cited in
eq.\ (\ref{Ustar}).

Subject to \emph{a posteriori} verification, assume the following trial
solutions for the first-order, state-specific local moments,
\begin{equation}
P_{\alpha}^{1}\approx P_{\alpha}^{0,\infty}\left(  \bar{U}^{\ast}t+B_{\alpha
}\right)  +\text{exp,}\label{Pa1}%
\end{equation}
where the $B_{\alpha}$ are time-independent state-specific constants to be
determined. \ Form the first moment of eq.\ (\ref{Pastate}) with $m=1$, and
use eq.\ (\ref{Painf}) for $\alpha=\left(  a,b\right)  $ together with the
trial solution (\ref{Pa1}) to obtain
\begin{equation}
\bar{U}^{\ast}-U_{\alpha}=kK_{\alpha}\left(  B_{\beta}-B_{\alpha}\right)
.\label{Beqn1}%
\end{equation}
As in prior macrotransport analyses \cite{Brenner:93}, the latter equation
defines each $B_{\alpha}$ only to within a common, state-independent arbitrary
constant, whose value proves irrelevant to the value of $\bar{D}^{\ast}$.
\ Substituting eq.\ (\ref{Ustar}) into (\ref{Beqn1}) furnishes the following
relation between the $B_{\alpha}$:
\begin{equation}
B_{a}-B_{b}=\frac{U_{a}-U_{b}}{k(K_{a}+K_{b})}.\label{B}%
\end{equation}
The latter time-independent solution, jointly with the fact that summing
eq.\ (\ref{Beqn1}) for $\alpha=\left(  a,b\right)  $ serves to reproduce
eq.\ (\ref{Ustar}), constitutes \emph{a posteriori} verification of the trial
solution (\ref{Pa1}). \ Substituting eqs.\ (\ref{Pa1}) for $\alpha=\left(
a,b\right)  $ into eq.\ (\ref{total}) with $m=1$ furnishes the first-order
total moment,
\begin{equation}
M_{1}\approx\bar{U}^{\ast}t+\frac{K_{b}B_{a}+K_{a}B_{b}}{K_{a}+K_{b}%
}+\text{exp},\label{M1}%
\end{equation}
which grows linearly in time at the rate $\bar{U}^{\ast}$. \

Substitution of eqs.\ (\ref{Painf}) and (\ref{Pa1}) for $\alpha=\left(
a,b\right)  $ into eq.\ (\ref{Meqn}) with $m=2$ yields
\begin{equation}
\frac{dM_{2}}{dt}\approx2\left(  \bar{U}^{\ast}\bar{U}^{\ast}t+\bar{D}%
^{M}+\frac{U_{a}K_{b}B_{a}+U_{b}K_{a}B_{b}}{K_{a}+K_{b}}\right)
+\text{exp},\label{dM2}%
\end{equation}
wherein $\bar{D}^{M}$ is given by eq.\ (\ref{Dm}). Use of the latter equation,
together with eqs.\ (\ref{Ddef}), (\ref{Ustar}), (\ref{B}) and (\ref{M1}),
furnishes the particle dispersivity (\ref{Dstar}).

\section*{Figure Captions}

\begin{description}
\item  Figure 1. Comparison between the numerical solution, $P$ (solid line),
of the microscale equations and the solution, $\bar{P}$ (dashed line), of the
macrotransport equation for the relatively short time $\tilde{t}=0.5$, and for
four different choices of the equation parameters: (a) $K = 1.5$, $\tilde{U} =
-2.0$, $\phi_{a} = 0.75$; (b) $K = 2.0$, $\tilde{U} = 0$, $\phi_{a} = 0.25$;
(c) $K = 0.75$, $\tilde{U} = 0.5$, $\phi_{a} = 0.33$; (d) $K = 0.5$,
$\tilde{U} = -0.5$, $\phi_{a} = 1$.

\item  Figure 2. Comparison between the numerical solution, $P$ (solid line),
of the microscale equations and the solution, $\bar{P}$ (dashed line), of the
macrotransport equation for the relatively long time $\tilde{t}=50$, and for
four different choices of the equation parameters: (a) $K = 1.5$, $\tilde{U} =
-2.0$, $\phi_{a} = 0.75$; (b) $K = 2.0$, $\tilde{U} = 0$, $\phi_{a} = 0.25$;
(c) $K = 0.75$, $\tilde{U} = 0.5$, $\phi_{a} = 0.33$; (d) $K = 0.5$,
$\tilde{U} = -0.5$, $\phi_{a} = 1$.
\end{description}

\begin{figure}
\includegraphics[clip,width=5in]{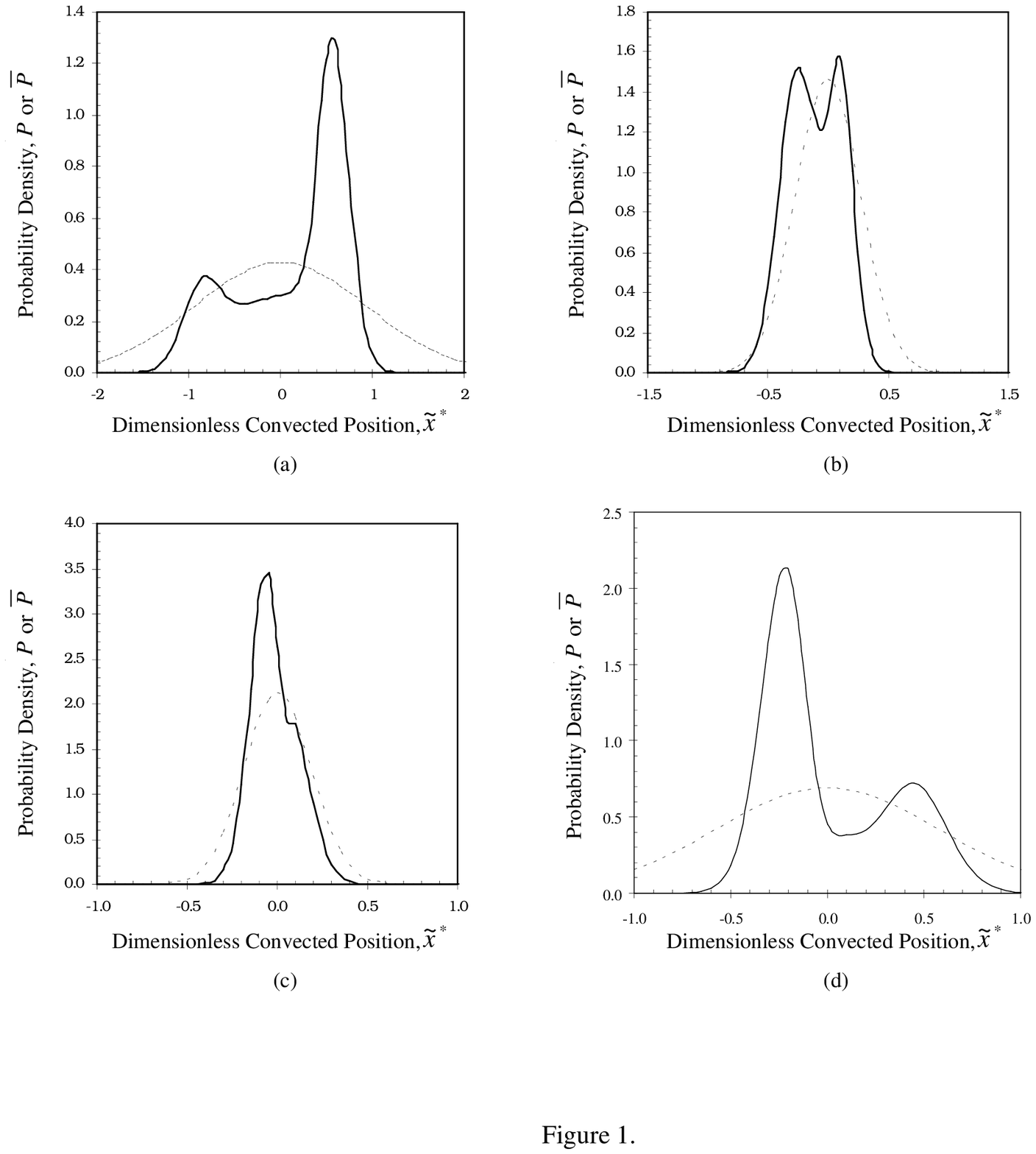}
\end{figure}

\begin{figure}
\includegraphics[clip,width=5in]{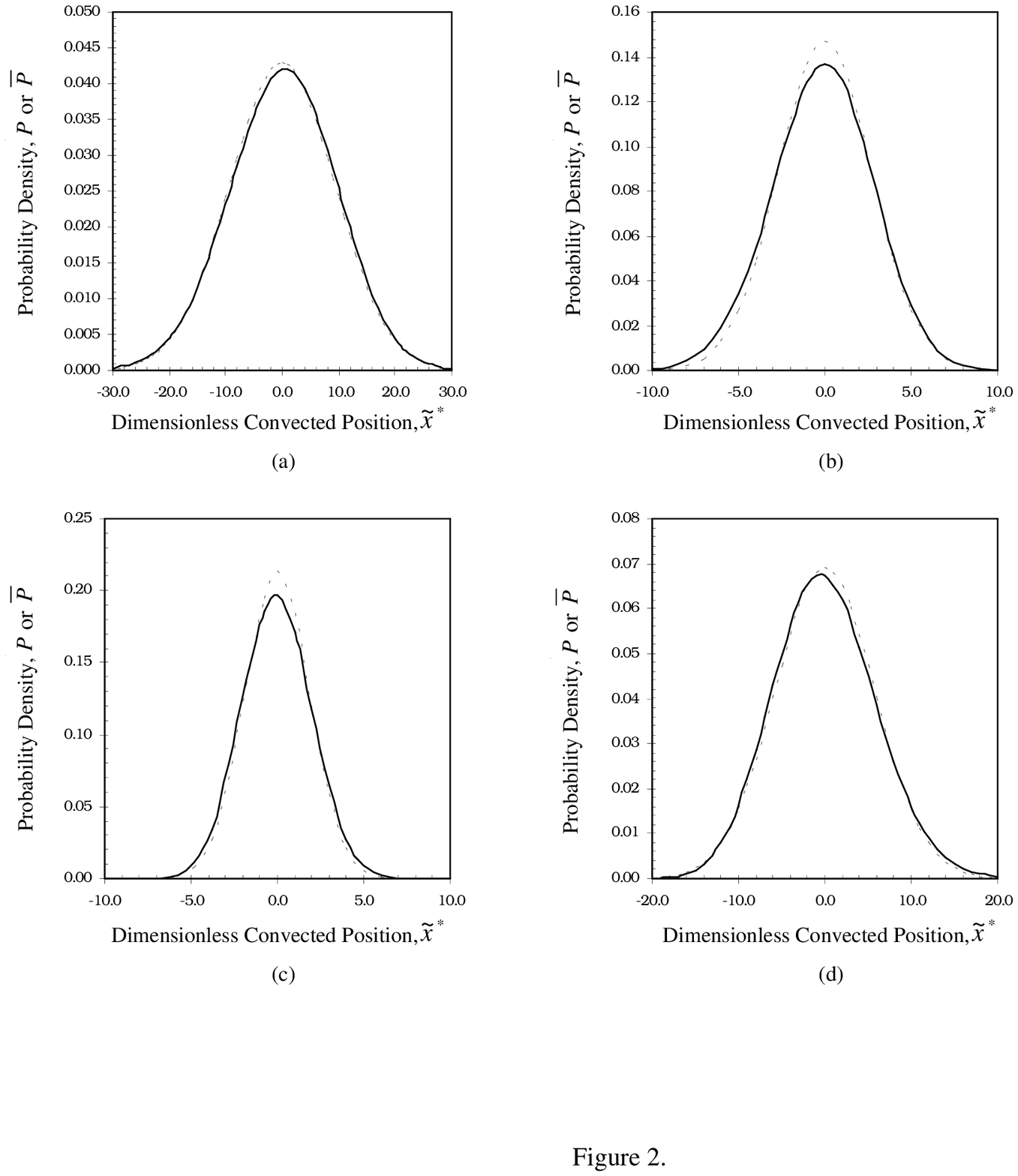}
\end{figure}


\begin{thebibliography}{99}
\bibitem{Giddings:55}J.~C. Giddings, H.~Eyring, A molecular dynamic theory of
chromatography, J.  Phys. Chem. 59 (1955) 416--421.

\bibitem{Balakotaiah:95}V.~Balakotaiah, H.~C.~Chang, Dispersion of chemical
solutes in chromatographs and  reactors, Phil. Trans. Roy. Soc. A 351 
(1995) 39--75.

\bibitem{Gitterman:95}M.~Gitterman, New applications of the two-state random
model, Physica A 221  (1995) 330--339.

\bibitem{Julicher:97}F.~Julicher, A.~Ajdari, J.~Prost, Modeling molecular
motors, Rev. Mod. Phys.  69~(4) (1997) 1269--1281.

\bibitem{Aris:59}R.~Aris, On the dispersion of a solute by diffusion,
convection and exchange  between phases, Proc. Roy. Soc. London A 252 
(1959) 67--77.

\bibitem{Brenner:93}H.~Brenner, D.~A.~Edwards, Macrotransport Processes,
Butterworth-Heinemann,  Boston, 1993.

\bibitem{new1}V.~N.~Kristensen, D.~Kelefiotis, T.~Kristensen, 
A.-L.~Borrensen-Dale,
High-throughput methods for detecting genetic variation, 
Biotechniques 30 (2001) 318--332.

\bibitem{new2}A.~J.~Nataraj, I.~Olivos-Glander, N.~Kusukawa, 
W.~E.~Highsmith~Jr., Single-strand conformation polymorphism and 
heteroduplex analysis for gel-based mutation detection, 
Electrophoresis 20 (1999) 1177-1185.

\bibitem{Taylor:53}G.~I.~Taylor, Dispersion of soluble matter in solvent
flowing slowly through a  tube, Proc. Roy. Soc. London A 219 (1953) 186--203.

\bibitem{McQuarrie:67}D.~A.~McQuarrie, Stochastic approach to chemical
kinetics, J. Appl. Prob. 4 (1967)  413--478.

\bibitem{Gillespie:77}D.~T.~Gillespie, Exact stochastic simulation of coupled
chemical reactions, J.  Phys. Chem. 81~(25) (1977) 2340--2361.

\bibitem{Pagitsas:86}M.~Pagitsas, A.~Nadim, H.~Brenner, Multiple time scale
analysis of  macrotransport processes, Physica A 135A (1986) 533--550.

\bibitem{Haber:93}S.~Haber, H.~Brenner, Effect of entrained colloidal
particles in enhancing the  transport of adsorbable chemical contaminants, J.
Colloid Interface Sci. 155  (1993) 226--246.

\bibitem{Iosilevskii:95}G.~Iosilevskii, H.~Brenner, Taylor dispersion in
discrete reactive mixtures,  Chem. Eng. Comm. 133 (1995) 53--91.

\bibitem{Aris:56}R.~Aris, On the dispersion of a solute in a fluid flowing
through a tube, Proc.  Roy. Soc. London A 235 (1956) 67--77.
\end{thebibliography}
\end{document}